# A Secure Land Record Management System using Blockchain Technology

Md. Samir Shahariar
*Department of Computer Science and Engineering (CSE)*
*Khulna University of Engineering & Technology (KUET)*
Khulna-9203, Bangladesh
shahariar1707108@stud.kuet.ac.bd

Pranta Banik
*Department of Computer Science and Engineering (CSE)*
*Khulna University of Engineering & Technology (KUET)*
Khulna-9203, Bangladesh
banik1707044@stud.kuet.ac.bd

Md. Ahsan Habib
*Department of Computer Science and Engineering (CSE)*
*Khulna University of Engineering & Technology (KUET)*
Khulna-9203, Bangladesh
mahabib@cse.kuet.ac.bd

*Abstract*— A land record (LR) contains very sensitive information related to land *e.g.* owner, buyer, etc. Currently, almost all over the world, the LR is maintained by different governmental offices and most of them maintain the LR with paper-based approach. Some of the works focus to digitalize the existing land record management system (LRMS) but with some security concerns. A blockchain-based LRMS can be effective enough to solve the existing issues. This paper proposes a blockchain-based LRMS that (*i*) digitalizes the existing paper-based system, (*ii*) ensures LR privacy using an asymmetric cryptosystem, (*iii*) preserves LR integrity, (*iv*) facilitates a platform for trading land through an advertising agency, and (*v*) accelerates the process of changing ownership that saves time significantly. Besides, this paper also proposes a new way of character to integer mapping named C2I table that reduces around 33% overhead of text to integer conversion compared to ASCII table. The experimental results, analyses, and comparisons indicate the effectiveness of the proposed LRMS over the state-of-the-art systems.

## I. Introduction

Land is an integral part of human life. In Bangladesh, a piece of land is identified by dag number *a.k.a.* khasra number and plot number in the rural area and urban area, respectively, whereas the land of right is defined by the khatiayan number. A land record (LR) contains land-related sensitive information along with its ownership [1]. A system in charge of managing and collecting the details of LR is known as a land record management system (LRMS). Currently, the LR is maintained by several government offices *e.g.* land records office, land revenue office, sub-registry office, etc. Most of the offices employed the paper-based approach. This strategy is high-risk and questionable. In contrast, very often people forge the LR and sell it to others.

In Bangladesh, the existing paper-based LRMS currently faces several limitations such as LRMS is not digitalized and a single LR is fragmented among multiple governmental offices. Some of the offices may be tainted in various ways, which allow them to simply alter the data from the records. Besides, LR is stored in a plain format that may violate the privacy of the landowner. In contrast, land trading is a very tedious and inefficient process. First, a buyer needs to confirm the ownership of the land to be purchased from the land office, then he checks and verifies bia deed, khatiayan, mutation, etc. of the property in favor of the seller. Then he surveys the land physically and prepares a deed of transfer. And finally, he applies for registration at the relevant sub-registry office presenting the receipts of payment of the registration and other fees. It is needed on average 2 months to finish the whole process [2].

Some of the existing works replace the paper-based LRMS with digitalized LRMS consisting of a centralized server. But, the systems are prone to different types of security attacks, alterations, manipulations, etc [3]. Hence, the systems remain still questionable. A blockchain-based LRMS can handle the existing issues efficiently and effectively. This paper proposes a blockchain-based LRMS that mitigates the existing concerns and accelerates the entire process. The contribution of this paper is referred as follows:

i. Digitalizing the traditional paper-based LRMS.
ii. Assuring the privacy of LR through encryption. An asymmetric cryptosystem is employed in this regard.
iii. Preserving the integrity of LR by employing blockchain technology that also provides the decentralization of the LRMS.
iv. Introducing a platform for trading land through an advertising agency. A landowner who wants to sell its land posts on the platform. A person who wants to buy quests for the desired land using the platform.
v. Accelerating of changing ownership of land which saves time significantly.
vi. Proposing a new approach of mapping character to integer named C2I table. By employing the C2I table in the conversion of text data to integer data around 33% overhead will be lessened compared to ASCII table.

The rest portions of the paper are organized as follows. Section II analyses the related works. Section III explains the required building blocks. Section IV explains the proposed C2I table, and section V demonstrates the proposed LRMS. Section VI describes the experimental results, and section VII concludes the paper.

## II. Related Works

Several existing works already have been proposed to address the limitations of the traditional LRMS.

The system in [4] proposed a three-stage blockchain-based land title management system for Bangladesh that enabled data synchronization, transparency, simplicity of access. It created a prototype system using Ethereum and gave a detailed design for smart contracts. Another system in [5] proposed a blockchain-based land registration system for Pakistan. It ensured trust and transparency of the system by employing a private blockchain and presented a conceptual framework only. However, both systems stored plain LR in the blockchain.

The system in [6] developed a land administration system and presented a smart contract covered specific use cases including sharing of ownership, transferring part of ownership, splitting or merging of real estate, and limiting the possibility of trading a real estate. Another framework in [7] also used blockchain technology and smart contract for LRMS. Similar to [4] and [5], both [6] and [7] did not encrypt LR before storing on the blockchain. Another scheme in [8] implemented a land registration using blockchain to maintain LRs easily and to secure the land transactions from attacker. It employed SHA256 algorithm to secure the LR transactions and Elliptic Curve Cryptography (ECC) for signatures. The scheme assessed under 12 nodes and 200 transactions only while employing Proof of Work (PoW) consensus mechanism which may consume very high resources.

The study in [9] designed blockchain-based for the implementation of tamper-proof, and provides authentic and conclusive rights on ownership land titling system with a focus on smart contracts. It introduced a platform where a purchaser guaranteed that the land being bought is the correct plot and that the seller is unambiguously the owner while lessening the disputes and forgery. Privacy of LR was not considered in the study. Another system [10] presented a smart contract application in the field of land administration where it focused on a quicker transaction execution process and eliminated double spending. No data encryption mechanism to protect the LR privacy was presented. A brief summary of the related works is presented in the following Table I.

TABLE I. BRIEF SUMMARY OF THE RELATED WORKS

| Feature | [4] | [5] | [6] | [7] | [8] | [9] | [10] |
|---|---|---|---|---|---|---|---|
| Store plain LR | √ | √ | √ | √ | × | √ | √ |
| Platform | E | PB | E | E | PB | E | E |

E: Ethereum; PB: Private Blockchain

Analyzing all the state-of-the-art works, there exists some research gaps which must be incorporated in a realistic LRMS. Firstly, almost all the works stored plain LR in the blockchain which could breach the data privacy. To the best of our knowledge, no system embeds any trading platform and that may lead the manual process again. And some of the systems take huge time in processing transactions and creating a new block which leads LRMS impractical. However, the proposed LRMS in this paper addresses the limitations.

III. REQUIRED BUILDING BLOCKS

A. Blockchain

Blockchain is a distributed ledger and tamper-proof information storage technology that keeps a historical record of all transactions that have taken place across a peer-to-peer network. The thought blockchain was first introduced by Satoshi Nakamoto in 2008 [11]. Since then, it has been widely used in different fields including public administration, supply chain management, healthcare, and many more [12]. The data is stored in blocks of blockchain that are cryptographically linked together to form a chain. Each block contains a cryptographic hash pointer of the previous block, a timestamp, and transaction data. The hash pointer links to the previous block and gives the immutability property of blockchain. New blocks are added only when the majority of nodes in the blockchain have agreed to it by validating all transaction data. As new blocks are added to the blockchain, its size continues to grow. A simplified blockchain structure is illustrated in the following Fig. 1.

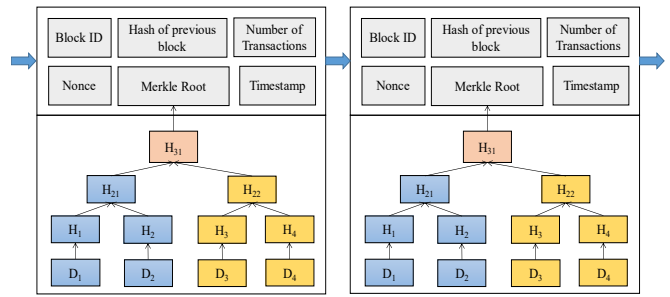

Fig. 1. Simple blockchain structure.

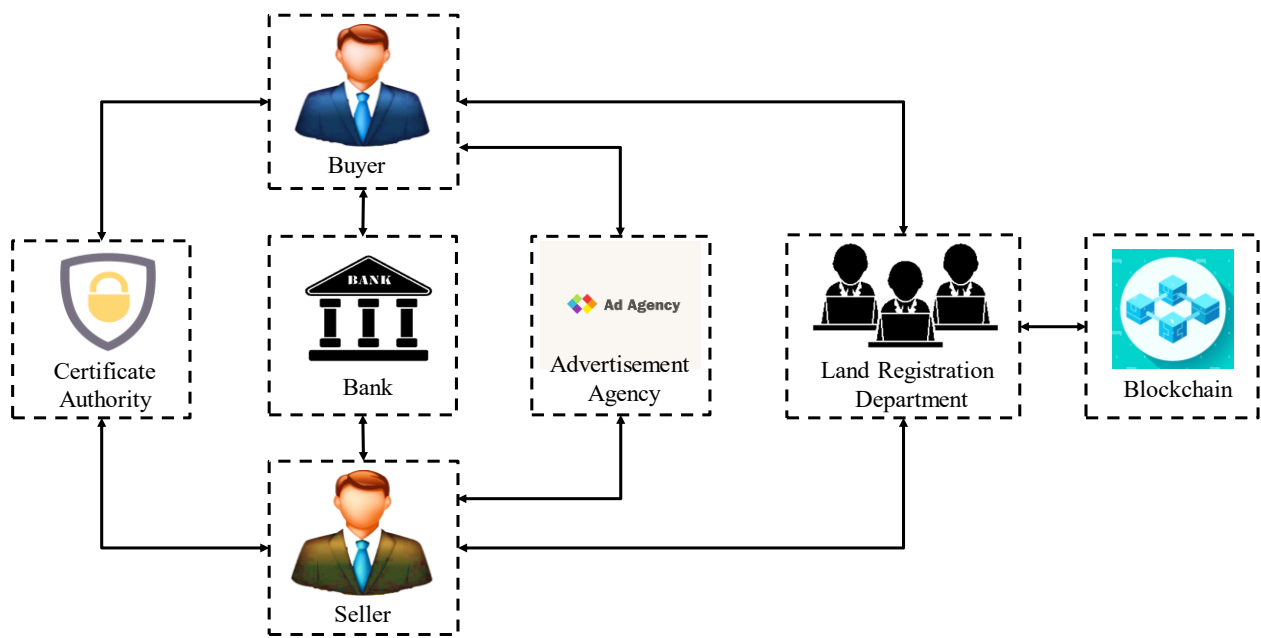

Fig. 2. The architecture of the proposed LRMS.

The block header and the block body are two parts of each block. The block header contains the metadata that typically includes block id, the hash of the previous block, number of transactions, nonce, Merkle root, timestamp, etc. whereas the block body contains each transaction data. The Merkle root is calculated from the binary hash tree called the Merkle tree. Here $D_i$ represents transaction data and $H_i$ denotes the cryptographic hash of transaction $D_i$. Secure Hashing Algorithm (SHA)-256 is a very prominent algorithm for hashing used in the blockchain domain.

*B. ElGamal encryption*

The ElGamal asymmetric key cryptosystem works using the following three stages (*KeyGen*, *Encryption*, *Decryption*). Here, a sender encrypts the data and send it to the receiver.

*KeyGen*: The receiver chooses a large prime number $p$ and a primitive element $\alpha$. Then it chooses a private key $a$ that satisfies $1 < a < p-2$ and compute $\beta = \alpha^a \bmod p$. Here, the $p$, $\alpha$ and $\beta$ remain public.

*Encryption*: Let the message is $x$ that satisfies $1 < x < p-1$. The sender chooses a random number $k$ that satisfies $1 < k < p-2$. Then it computes,

$$y_1 = \alpha^k \bmod p$$
$$y_2 = x * \alpha^k \bmod p$$

*Decryption:* The receiver retrieves the plain message $x$ by computing $D$.

$$D = \frac{y_2}{y_1^a} \bmod p$$
$$= \frac{x * \beta^k \bmod p}{\alpha^{ka} \bmod p} \bmod p$$
$$= \frac{x * \alpha^{ak}}{\alpha^{ak}} \bmod p$$
$$= x \bmod p$$
$$= x \text{ [Because } x < p\text{]}$$

## IV. PROPOSED C2I TABLE

A new character to integer mapping approach named C2I table is presented in this paper. The motivation behind developing the C2I table is the ASCII table requires 3 decimal digits when converting a single character to its corresponding integer value. There are 256 individual characters presented in the ASCII table, most of them remains useless most of the time. The C2I table considers 95 individual characters that are usually useful. As there are only 95 characters in C2I table, it is needed only 2 decimal digits when converting a single character to its integer. It clearly shows that, by employing the C2I table in the conversion of text data to integer data around 33% overhead will be lessen compared to the ASCII table.

## V. PROPOSED LRMS

The overall architecture of the proposed blockchain-based LRMS is demonstrated in Fig. 2 and explained below.

*A. Involved Entities*

- *Certification Authority (CA):* An entity that acts to verify the identities of entities and bind them to cryptographic keys known as digital certificates is referred to as a *CA*. Both the seller and the buyer get digital certificates from the *CA*.
- *Buyer (B):* A person who wants to purchase land. *B* chooses the land and negotiates a deal with the seller.
- *Seller (S):* An individual who owns the land and wants to sell it.
- *Advertisement Agency (AA):* A platform that is used for trading lands. *S* posts the land it wants to sell on this platform and *B* searches for the required land here.
- *Bank (Bn):* An organization that acts as an intermediary between *B* and *S* for a transaction and offers loans also.
- *Land Registration Department (LRD):* A government organization that manages land-related information and stores it on the private blockchain after encryption.
- *Blockchain Storage Server (BSS):* The encrypted data is kept on it by *LRD*.

*B. Working Principle of the Proposed LRMS*

Assuming the *LRD* already created and stored the existing land details after encryption for every landowner. The working procedure of the proposed LRMS as follows.

TABLE II. CHARACTER TO INTEGER (C2I) MAPPING

| Character | Integer | Character | Integer | Character | Integer | Character | Integer | Character | Integer | Character | Integer |
|---|---|---|---|---|---|---|---|---|---|---|---|
| 0 | 01 | H | 18 | Y | 35 | p | 52 | ' | 69 | ^ | 86 |
| 1 | 02 | I | 19 | Z | 36 | q | 53 | ( | 70 | _ | 87 |
| 2 | 03 | J | 20 | a | 37 | r | 54 | ) | 71 | [ | 88 |
| 3 | 04 | K | 21 | b | 38 | s | 55 | * | 72 | \ | 89 |
| 4 | 05 | L | 22 | c | 39 | t | 56 | + | 73 | ] | 90 |
| 5 | 06 | M | 23 | d | 40 | u | 57 | , | 74 | ` | 91 |
| 6 | 07 | N | 24 | e | 41 | v | 58 | - | 75 | ~ | 92 |
| 7 | 08 | O | 25 | f | 42 | w | 59 | . | 76 | { | 93 |
| 8 | 09 | P | 26 | g | 43 | x | 60 | / | 77 | \| | 94 |
| 9 | 10 | Q | 27 | h | 44 | y | 61 | : | 78 | } | 95 |
| A | 11 | R | 28 | i | 45 | z | 62 | ; | 79 | | |
| B | 12 | S | 29 | j | 46 | | 63 | < | 80 | | |
| C | 13 | T | 30 | k | 47 | ! | 64 | = | 81 | | |
| D | 14 | U | 31 | l | 48 | " | 65 | > | 82 | | |
| E | 15 | V | 32 | m | 49 | # | 66 | ? | 83 | | |
| F | 16 | W | 33 | n | 50 | % | 67 | @ | 84 | | |
| G | 17 | X | 34 | o | 51 | & | 68 | $ | 85 | | |

1. *S* and *B* both get digital certificates from the *CA*. The interaction is shown in the following Fig. 3.

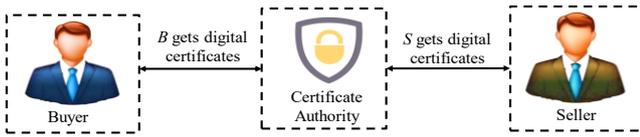

Fig. 3. Interaction between *CA* & *B* and *CA* & *S*.

2. *S* login to the *LRD* website to confirm its identity. Then *S* asks its land-related information.

3. Then the *LRD* retrieves the required information from *BSS* and sends back to the *S*. The communications are shown in the following Fig. 4.

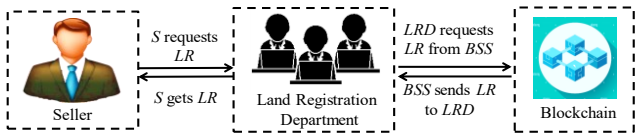

Fig. 4. Interaction between *LRD* & *S* and *LRD* & *B*.

4. *S* posts the land details on *AA* for sale. *B* selects its preferred land from *AA* to buy.

5. *B* login to the *LRD* website to verify its identity.

6. *AA* creates a deed and sends it to both the *B* and *S*. A sample deed is shown in Fig. 5. The interactions are depicted in the following Fig. 6.

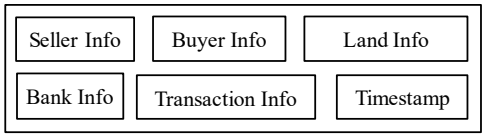

Fig. 5. Interaction between *AA* & *S* and *AA* & *B*.

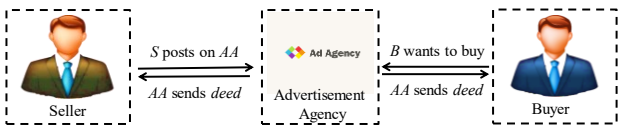

Fig. 6. Interaction between *AA* & *S* and *AA* & *B*.

7. Both the *B* and *S* sign the deed and send it to the *Bn* for further procedure. The *Bn* signs and returns the deed to both the *B* and *S* after successful transaction. The contacts are illustrated in the following Fig. 7.

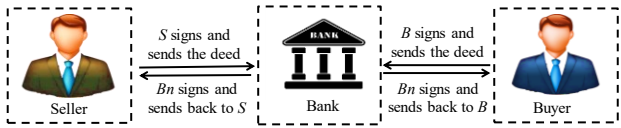

Fig. 7. Interaction between *Bn* & *S* and *Bn* & *B*.

8. The deed is sent to the *LRD* for updating the ownership details.

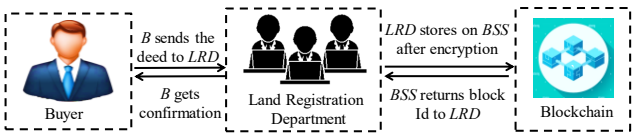

Fig. 8. Interaction between *B* & *LRD* and *LRD* & *BSS*.

9. The *LRD* creates a new block with the updated information and append it to the blockchain. Here, in Fig. 8 the interactions are demonstrated.

C. *Encryption Phase*

The *LRD* encrypts the land records by using the public key of its owner $Pub_{owner}$ through the following steps. The operation flow of the encryption process is depicted in the following Fig. 9.

Step 1: First, read the *LR* which usually contains text data.

Step 2: Convert the text data into the corresponding integer value based on C2I table.

Step 3: Encrypt the integer data by using an asymmetric cryptosystem.

Step 4: Convert the encrypted data into binary bits.

Step 5: Transform bits into corresponding DNA bases (11 = A, 10 = T, 01 = C, 00 = G) and store encrypted data in the *BSS*.

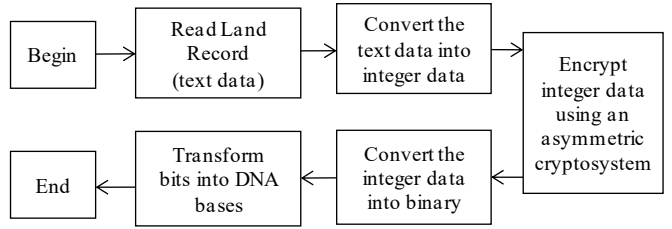

Fig. 9. Encryption process.

D. *Blockchain Storage Phase*

After encryption of *LR*, the *LRD* creates a new block on the *BSS*. The *BSS* returns the corresponding block id $Id_{block}$ to the *LRD* for further usage. Then the *LRD* stores the $Id_{block}$ along with the identity of its corresponding owner.

E. *Retrieval Phase*

If a landowner wants to retrieve its *LR*, it requests to the *LRD*. Then the *LRD* requests the specific block of landowner to *BSS* by providing a $Id_{block}$ and it returns the corresponding block to the *LRD*. Then the *LRD* sends back the *LR* to its owner. Upon receiving the encrypted *LR* from *LRD*, it decrypts the encrypted *LR* using the private key $Pri_{owner}$. The decryption steps are mentioned in the followings. The operation flow of the retrieval process is depicted in Fig. 10.

Step 1: Read encrypted data from the *BSS*.

Step 2: Decode the DNA encoded data to retrieve binary bits.

Step 3: Convert the binary bits into corresponding integer data.

Step 4: Decrypt the individual integer data using the private key.

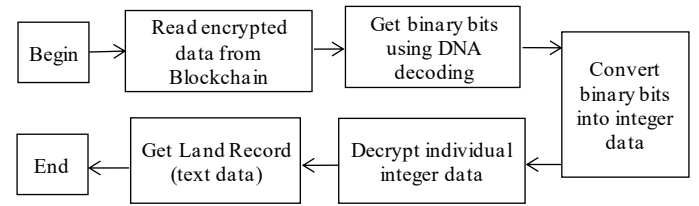

Fig. 10. Retrieval process.

TABLE III. OUTPUT OF THE ENCRYPTION PHASE

| Step | Operation | Output |
|---|---|---|
| Step 1 | Read the land record (Text data) | Seller: Mr. X, Buyer: Mr. Y, Land information: Dag number: 8000, Khatiayan number: 3000, Area:2000 Shotangsho, Transaction ID: BNXY2345 |
| Step 2 | Convert the text data into the corresponding integer value | 29414848415478632354766374633463125761415478632354766335746322375040631950425* |
| Step 3 | Encrypt the text data using ElGamal | 0800207809710910971111110130640642440803206510810509910116001024016601618617101719113901923917902919911865449375612301318301455150660310561051 0* |
| Step 4 | Convert the encrypted data into bits | 1000011001100100011101011101011111101101100101111110101000001011100111001010011101101011101011000100010011001011111001101110010100001001111 10* |
| Step 5 | Transform bits into corresponding DNA bases | TGCTCTCGCACCACCAATACTCCAAACCGGCCAGATCCGATACCACCTGTGTCTCCAAG ACAGTTGCGAATTGCTCTCGCACCACCAATACTCCAAACCGGCCAGATCCGATA* |

*denotes a portion of data

TABLE IV. OUTPUT OF THE RETRIEVAL PHASE

| Step | Operation | Output |
|---|---|---|
| Step 1 | Read encrypted data from blockchain | TGCTCTCGCACCACCAATACTCCAAACCGGCCAGATCCGATACCACCTGTGTCTCCAAG ACAGTTGCGAATTGCTCTCGCACCACCAATACTCCAAACCGGCCAGATCCGATA* |
| Step 2 | Get binary bits using DNA decoding | 1000011001100100011101011101011111101101100101111110101000001011100111001010011101101011101011000100010011001011111001101110010100001001111 10* |
| Step 3 | Convert binary bits into integer data | 0800207809710910971111110130640642440803206510810509910116001024016601618617101719113901923917902919911865449375612301318301455150660310561051 0* |
| Step 4 | Convert the encrypted data into bits | 29414848415478632354766374633463125761415478632354766335746322375040631950425* |
| Step 5 | Transform bits into corresponding DNA bases | Seller: Mr. X, Buyer: Mr. Y, Land information: Dag number: 8000, Khatiayan number: 3000, Area:2000 Shotangsho, Transaction ID: BNXY2345 |

*denotes a portion of data

Step 5: Retrieve the land record (text data) from the integer data using the C2I table.

## VI. EXPERIMENTAL STUDIES

### A. Experimental Setup

The prototype of the proposed technique is developed under the environment on Intel(R) Core™ i5-7300HQ CPU @2.50 GHz 64-bit processor with 8GB of RAM running on Windows 10 OS. It was developed in VS code 2019. The ElGamal cryptosystem is adopted as an asymmetric cryptosystem and used a 1024-bit length key for operations.

### B. Output of the Encryption Phase

Considering the plain LR 'Seller: Mr. X, Buyer: Mr. Y, Land information: Dag number: 8000, Khatiayan number: 3000, Area:2000 Shotangsho, Transaction ID: BNXY2345' and using the steps of the section IV(C), Table III depicts the output of the encryption phase.

### C. Output of the Decryption Phase

Using the steps of the section IV(E) and the LR of section V(B), Table IV presents the output of the decryption phase.

### D. Comparison with other LRMS

Table V demonstrates the suitability of the proposed LRMS for data privacy and authenticity over the state-of-the-art works. In summary, the proposed LRMS ensures LR privacy and provides users' authenticity where most of the systems fails to do. By incorporating private blockchain, the system offers very low cost (in fact free if government want). The proposed LRMS facilitates a platform for trading land through an advertising agency. Additionally, the system enables quick processing times for usage by using C2I table.

TABLE V. COMPARISON WITH OTHER SYSTEMS

| Feature | [4] | [5] | [6] | [7] | [8] | [9] | [10] | Proposed |
|---|---|---|---|---|---|---|---|---|
| Data Privacy | × | × | × | × | √ | × | × | √ |
| Authenticity | − | − | − | − | − | − | − | √ |
| Cost | H | L | H | H | L | H | H | L |
| Trading Platform | × | × | × | × | × | × | × | √ |

×: not applied; √: applied; −: not mentioned; H: High; L: Low

### E. Security and Performance Analysis

This section assesses the security aspects, *i.e.*, data privacy, data integrity, etc., encompassed by the proposed system.

*Data Privacy:* The proposed scheme offers privacy via ElGamal encryption. Because of the robustness and probabilistic encryption over other asymmetric cryptosystems, ElGamal is chosen.

*Data Integrity:* As the blockchain is an immutable ledger, data integrity is ensured by storing in a distinct block of the blockchain.

*Time:* By employing the C2I table, around 33% conversion time will be lessened compared to other scheme.

## VII. DISCUSSION AND CONCLUSION

This paper proposes a blockchain-based land record management system to secure land record management system maintains adequate data privacy, data integrity, availability, etc., about the land record. Herein, an asymmetric cryptosystem is exploited to enrich data privacy. The storage of the land record over the blockchain assures integrity. The trading platform alleviates the burden of searching the proper land. The proposed C2I table reduces on average 33% overhead of text to integer conversion compared to ASCII table. The preliminary assessment refers the effectiveness of the proposed scheme. A forthcoming plan is to improve the encryption and decryption process, and implement the prototype of the proposed land record management system in a more realistic environment.